\documentstyle[graphicx,twocolumn,aps]{revtex}
\draft
\preprint{LA-UR-99-5452}

\newcommand{\etal}      {{\it et al}}
\newcommand{\cuot}      {CuO$_2$}
\newcommand{\lesco}     {${\rm La}_{1.65} {\rm Eu}_{0.2} {\rm Sr}_{0.15} {\rm Cu O_4}$}
\newcommand{\lescox}    {${\rm La}_{1.8-x} {\rm Eu}_{0.2} {\rm Sr}_{x} {\rm Cu O_4}$}

\newcommand{\lsco}      {${\rm La}_{1.85} {\rm Sr}_{0.15} {\rm Cu O_4}$}
\newcommand{\lscox}     {${\rm La}_{2-x} {\rm Sr}_{x} {\rm Cu O_4}$}
\newcommand{\lc}        {lanthanum cuprate}

\newcommand{\lilco}     {$\rm La_2 Cu_{1-x} Li_{x} O_{4}$}

\newcommand{\la}        {$^{139}$La}

\newcommand{\ltoi}      {$^{139}T_1^{-1}$}
\newcommand{\ctoi}      {$^{63}T_1^{-1}$}

\title{ Inhomogeneous Low Frequency Spin Dynamics in
\mbox{La$_{1.65}$Eu$_{0.2}$Sr$_{0.15}$CuO$_4$}}
  \author{N.J. Curro,$^{1}$ P.C. Hammel,$^{1}$ B.J. Suh,$^{1,2}$ M. H\"ucker,$^{3}$
  B. B\"uchner,$^{3}$ U. Ammerahl,$^{3,4}$ and A. Revcolevschi$^{4}$}
  \address{$^{1}$Condensed Matter and Thermal Physics, Los Alamos
  National Laboratory, Los Alamos, NM 87545, USA}
  \address{$^{2}$  Division of Natural Science, The Catholic University of Korea,  Korea}
  \address{$^{3}$II. Physikalisches Institut, Universit\"at zu K\"oln, Germany}
  \address{$^{4}$Laboratoire de Chimie des Solides,
  Universit\'e Paris-Sud, 91405 Orsay Cedex, France
  \vspace{-4mm}}
  \author{\small (Received: November 22, 1999)}

\address{ \parbox{14cm}{\bigskip\rm\small
We report Cu and La
nuclear magnetic resonance (NMR) measurements in the title
compound that reveal an inhomogeneous glassy behavior of the spin
dynamics. A low temperature peak in the La spin lattice relaxation
rate and the ``wipeout'' of Cu intensity both arise from these
slow electronic spin fluctuations that reveal a distribution of
activation energies. Inhomogeneous slowing of spin fluctuations
appears to be a general feature of doped lanthanum cuprate.
  }}

\begin{document} \maketitle

\thispagestyle{myheadings} \markright{{\em LA-UR-99-5452}}

\narrowtext

Lanthanum cuprate, the prototypical single layer high temperature
superconductor, has been extensively studied for several years to
understand the origin of its unusual normal state behavior as well
as the mechanism for superconductivity. Rare earth co-doped \lc\
has received attention recently because elastic neutron scattering
experiments have revealed ordering of doped holes into charged
stripes that constitute anti-phase domain walls producing
incommensurate antiferromagnetic (AF) order in the intervening
undoped domains\cite{tranqsuplat}.  Charge stripe order is likely
intimately related to the high temperature
superconductivity\cite{emkiv,neto,zaanen,scalapino}. Isostructural
lanthanum nickelate demonstrates clear stripe
order\cite{swc:NSlsno}, and it has been shown there that both the
charge order and the magnetic order are
glassy\cite{swc:NSlsno,yosh:lsno}. It is also known that the
magnetic order associated with charge ordering in \lc\ is
glassy\cite{tranqglass,julien:glass}, but this situation is more
difficult because the charge superlattice peaks are very hard to
observe, presumably because the stripes tend to be dynamic.  As a
consequence little detail is known about the glassy behavior. Hunt
\etal . have observed suppression of the Cu NQR signal intensity
(``wipeout'') with decreasing temperature that they attribute to
charge stripe order\cite{hunt}.

NMR provides information complementary to neutron scattering
because the nuclei are sensitive to the local magnetic field and
the dynamic behavior of the electronic system without requiring
spatial correlations.  Chou {\it et al.,} first proposed that the
very strong peak in the \la\ nuclear spin relaxation rate \ltoi\
in the vicinity of $T \! = \! 10$ K is associated with spin
freezing\cite{chou}. Kataev \etal . have observed slow spin
fluctuations in \lescox\ at low $T$\cite{kataev}.  Our single
crystal studies of \ltoi\ have shown that the peak is due to
continuously slowing electronic spin fluctuations; in particular
the characteristic fluctuation frequency $\tau_c^{-1}$ displays an
activated temperature dependence\cite{suhlesco}. Furthermore,
these data demonstrate a distribution $P(E_a)$ of activation
energies $E_a$ centered at $E_a/k_B T \sim 50 $ K and with a width
comparable this center value indicating strongly inhomogeneous
magnetic properties\cite{suhlesco}. To understand if this
inhomogeneity arises from disorder due to, e.g., substitutional
dopants, we have applied this analysis to several \lc\ systems
exhibiting AF order at low temperatures to allow us to explore the
effect of varying the density and character of the disorder:
in-plane doping by Li substitution for Cu, variation of doping
density in LTT phase \lescox : $0.01 \le x \le 0.15$. Remarkably,
we find that the character of the inhomogeneity, that is, the
distribution of activation energies is essentially unchanged in
all these cases and very similar to lightly doped
\lscox\cite{chou}, suggesting that this inhomogeneity is {\it
intrinsic} rather than arising from impurity disorder.

These inhomogeneous slow spin fluctuations also enhance the Cu
spin relaxation rate \ctoi and directly explain the wipeout of Cu
intensity reported by Hunt \etal .\cite{hunt}. Because Cu nuclear
moments experience a hyperfine coupling $A_{\rm hf}$ to these
fluctuations that is two orders of magnitude larger than for \la\
and $T_1^{-1} \propto A_{\rm hf}^2$, \ctoi\ becomes so fast as to
relax the signal before it can be observed. The distribution of
electron spin fluctuation frequencies means that different Cu
spins will move out of the observation window of the spectrometer
at different temperatures; using $P(E_a)$ obtained from \ltoi\ we
quantitatively explain the loss of Cu intensity.

Co-doping with $J\!=\!0$ Eu rather than Nd is advantageous:
neither the magnetism in the CuO$_{2}$ plane nor the nuclear
magnetism suffer the effects of the large Nd magnetic moment. The
crystal used in this study was grown using the traveling solvent
floating zone method under oxygen pressure of 3
bar\cite{buchnerxtal}. Diffraction data as well as La NMR indicate
a sharp LTT structural phase transition at $135 \pm 2$ K. The
observation of incommensurate magnetic peaks near 30 K in Eu
co-doped compounds by elastic neutron scattering\cite{NSlesco}
reveals static magnetic stripe order. From dc magnetization
measurements, no superconducting transition is observed down to
4.2K. $\mu$SR studies of this sample reveal static spin order
below 25K\cite{musrlesco}.

The $^{139}$La ($I=\frac{7}{2}$) and $^{63}$Cu ($I=\frac{3}{2}$)
NMR measurements were made on the central
($m_I=+\frac{1}{2}\leftrightarrow-\frac{1}{2}$) tran-
  \begin{figure}[t]
  \centering
  \includegraphics[width=0.9\linewidth]{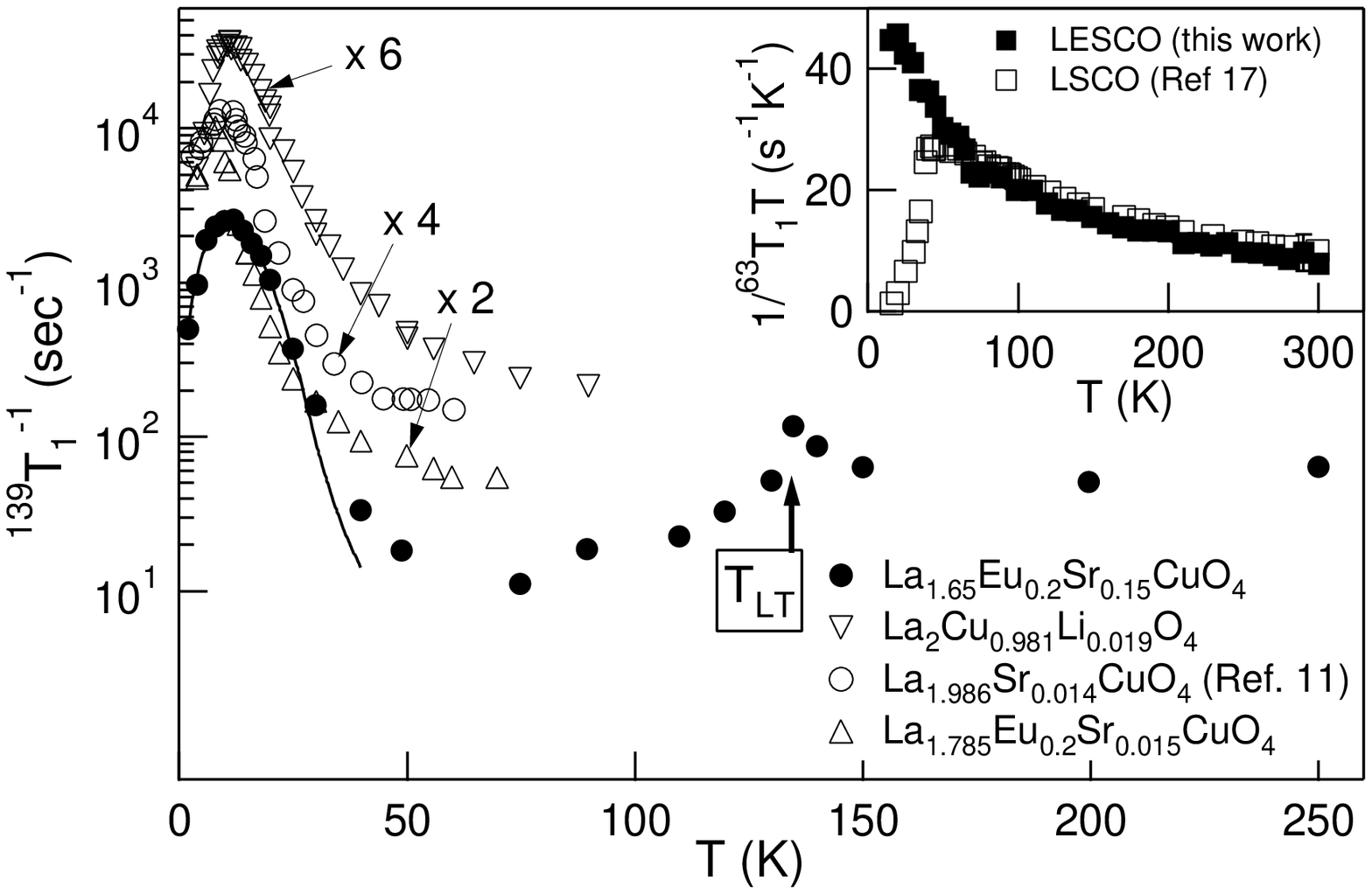}
  \caption{$T$ dependencies of $^{139}T_1^{-1}$ in \lc\ doped by
  various routes ($\rm La_{1.986} Sr_{0.014} Cu O_4$
  ({\large$\circ$}) from \protect\cite{chou}). Solid lines are
  fits as described in the text using the
  parameters shown in Table 1.
  Inset: $(^{63}T_1T)^{-1}$ versus $T$ in \lesco\ (solid squares)
  and \lsco\ (open squares; from \protect\cite{ohsugi}).}
  \label{fig:lat1}
  \end{figure}
  \noindent
sition. The spin lattice relaxation rates were measured by
monitoring the recovery of the magnetization after an inversion
pulse, and the Cu echo decay was observed by integrating the spin
echo. The Cu spin lattice relaxation and echo decay data were
measured in a field of 84.2 kG, and the La spin lattice relaxation
data were obtained in a field of 59.7 kG. The intensities of the
Cu and La signals were obtained from field swept spectra at
constant frequency; the areas $A$ under the spectra were then
adjusted for echo decay and spin lattice relaxation effects.

\ltoi\ is ideal for probing slow electron spin fluctuations
because La is located outside of the CuO$_2$ planes, and so only
weakly coupled to the electronic system. For $T \gtrsim 50$ K,
except for a range of 35 K around $T_{LT}$, the magnetization
recovery is well fit by the standard expression for magnetic
relaxation of the central transition of a spin 7/2 nucleus with a
single component $T_{1}$\cite{relaxpaper}. Between 125 K and 160 K
and for $T \lesssim 50$ K, the magnetization recovery was fit with
the stretched exponential form:
 \(M(t)=M_0[1-2 \exp (-\sqrt {t/T_1})]\),
 % $M(t)=M_0[1-2e^{-(t/T_1)^{1/2}}]$,
where $M_0$ is the equilibrium magnetization. This expression
represents the magnetization recovery for a distribution of
relaxation rates, with a peak at $1/6T_1$. The data for \ltoi\ are
shown in Fig.\ \ref{fig:lat1}. The peak at 135 K reflects the LTT
transition, where the spin lattice relaxation is dominated by
quadrupolar components\cite{suhSPT}.

Recently, Suh and coworkers\cite{suhlesco} demonstrated that the
strong low temperature peak in ${}^{139}T_1^{-1}$ is accurately
described by the Bloembergen, Purcell and Pound
  \begin{table}
  \caption{Parameters describing distribution of activation energies:
  \(P(E_a)={\mathcal N} \exp[-(E_a-E_0)^2/2 \Delta ^2]\).
  Fits (see Fig.~\protect\ref{fig:lat1})
  were performed with fixed $\tau_{\infty}$=0.03 ps.}
  \begin{tabular}{lccc}
    \multicolumn{1}{c}{Material} & Ref. & $E_0/k_B$ (K) & $\Delta/k_B$ (K) \\ \hline
    \lesco & &73 & 84 \\
    $\rm La_2 Cu_{0.981} Li_{0.019} O_{4}$ &\protect\cite{suhllco}& 119 & 64 \\
    $\rm La_{1.986} Sr_{0.014} Cu O_4$ &\protect\cite{chou} & 62 & 80 \\
    $\rm La_{1.785} Eu_{0.2} Sr_{0.015} Cu O_4$ &\protect\cite{suhlesco}& 75 & 38 \\
  \end{tabular}
  \end{table}
\noindent mechanism (BPP)\cite{BPP} introduced to explain nuclear
spin relaxation that results when the characteristic electron spin
fluctuation frequency ($\tau_c^{-1}$) decreases continuously with
decreasing temperature: \(\tau_c=\tau_{\infty}\exp(E_a/k_BT)\).
The relaxation rate is given by the spectral density of
fluctuations (typically a Lorentzian) evaluated at the Larmor
frequency $\omega_L$\cite{CPS}: \begin{equation}
%  \frac{1}{T_{1}}=\gamma^2 h_{0}^2 \frac{\tau_c}{1+\omega_L^2\tau_c^2},
  1 / T_{1} = \gamma^2 h_{0}^2 \tau_c / (1+\omega_L^2\tau_c^2),
\label{eqn:t1vstau}
\end{equation}
where $h_{0}$ is the local field fluctuating at the nuclear site,
and $\gamma$ is the gyromagnetic ratio.  Hence, the peak occurs at
the temperature where the continuously slowing characteristic
frequency matches the measurement frequency: $\tau_c^{-1} \! = \!
\omega_L$. Since the peak temperature is inherently
probe-frequency dependent the dynamics are best described by
$E_a$.  A crucial difference from the standard BPP model is that a
distribution of $E_a$ is required to describe the relaxation data.
Roughly speaking, the high temperature side of the peak in \ltoi\
determines the center of the distribution, while the slow decrease
of \ltoi on low temperature side can only be explained if some
fraction of the sample experiences much smaller values of $E_a$.
We choose a Gaussian as a convenient distribution of activation
energies: \(P(E_a)={\mathcal N} \exp[-(E_a-E_0)^2/2 \Delta ^2 ]
\), where ${\mathcal N}$ is a normalization factor; we now
consider \ltoi\ to be a function of position describing relaxation
in some region of space at a particular time. This gives rise to a
distribution of spin lattice relaxation rates \(
P(^{139}T_{1}^{-1}) d^{139}T_{1}^{-1} = P(E_a)dE_a \). The
measured \ltoi\ is then given by: \( ^{139}T_1^{-1}=
\int_0^{\infty}\, ^{139}T_1^{-1} P(^{139}T_1^{-1})\,
d^{139}T_1^{-1} \). The solid line in Fig.\ \ref{fig:lat1} is a
fit to this expression.

To explore the dependence of this inhomogeneous distribution of
magnetic properties on impurity disorder we show \ltoi\ results
from several systems in Fig.\ \ref{fig:lat1}:  the same Eu-doped
cuprate with an order of magnitude lower hole doping: $\rm
La_{1.785} Eu_{0.2} Sr_{0.015} Cu O_4$\cite{suhSPT}; lightly
in-plane doped $\rm La_2 Cu_{.98} Li_{.02} O_{4}$\cite{suhllco};
and the Chou results\cite{chou}: $\rm La_{1.982} Sr_{0.018} Cu
O_4$. The evident similarity of the low temperature relaxation
data is confirmed by the similar distributions of $E_a$ (see Table
1). This indicates that impurity disorder is not the crucial
element for this inhomogeneity, rather it appears intrinsic.

\begin{figure}
  \centering
  \includegraphics[width=0.8\linewidth]{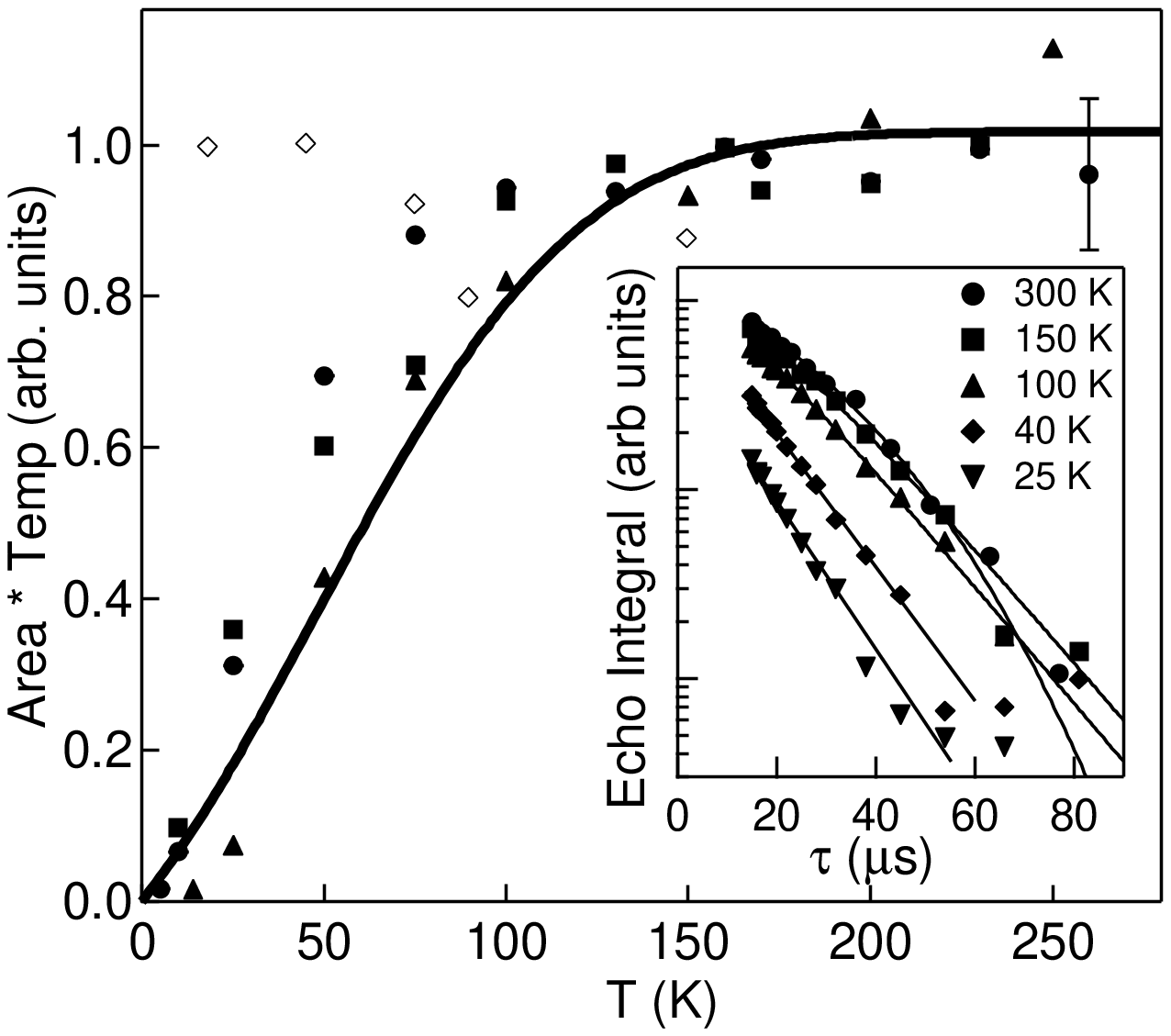}
  \caption{NMR measurements in \lesco\ showing the wipeout of Cu
  signal. The solid symbols represent Cu data: (squares) 77 MHz,
  $H_0 \! \perp \! c$; (circles) 86 MHz, $H_0 \! \perp \! c$;
  (triangles) 95 MHz, $H_0 ||c$.  The open diamonds represent La
  data for $H_0 ||c$ at 35.9 MHz. The solid line is a plot of
  $I_0(T)$ as described in the text with $\kappa$=1.5. Inset: $H_0
  || c \, ^{63}\!$Cu echo size versus the pulse spacing $\tau$ is
  plotted for a series of temperatures. The fits (solid lines) are
  described in the text.}
  \label{fig:cuint}
  \end{figure}

We observe the same wipeout of Cu signal intensity (see Fig.\
\ref{fig:cuint}) discussed extensively by Hunt \etal .\cite{hunt}.
$A$ is proportional to $N/T$, where $N$ is the number of nuclei,
and $T$ is the temperature, therefore $N\sim AT$ is proportional
to the number of nuclei which give rise to the NMR signal.  In
Fig.\ \ref{fig:cuint} we show $N(T)$ for Cu and La in various
fields and orientations in \lesco. About 96\% of the signal from
the Cu nuclei vanishes between 90 and 4 K, whereas none of the La
signal is lost. The Curie-Weiss broadening of the spectra is
typical for planar Cu in cuprates, however the spectra lose
intensity uniformly, and develop no anomalous features. An
apparent loss of signal due to a shift of intensity to another
region of the spectrum can be ruled out.  A field dependent study
  \noindent
revealed no evidence for a component with a different magnetic
shift. Even a large change in the local NQR frequency would be
easily detected in the frequency of the NMR central transition,
where the quadrupolar shift is second order (i.e., a 10 MHz change
in $\nu _Q$ would shift the central transition by $\sim$ 1--2
MHz). However, the temperature dependence of the $H_0\perp c$
quadrupolar shift with $H_0 \sim 80$ kG reveals no change in
$\nu_Q$ with temperature down to 4 K. Therefore the evidence
suggests that the loss of intensity is not due to an inhomogeneity
in either the local magnetic shift or the electric field gradient.

Rather, the slow spin fluctuations responsible for the strong low
temperature peak in \ltoi\ straightforwardly explains the
intensity data if we realize that the hyperfine field $h_0$ of the
electronic moments is two orders of magnitude larger at the Cu
site than it is at the La site (see Eq.\ \ref{eqn:t1vstau}). Thus
for nuclei experiencing slow fluctuations, $\tau_c> \tau_\infty
e^\kappa$ the nuclei relax so fast the spin echo signal has
decayed before it can be measured; $\kappa$ is a constant
determined by the recovery time $t_r \simeq 15\,\mu$sec of the
spectrometer following a spin-echo excitation pulse. The echo
decay of a Cu nucleus experiencing a particular local fluctuation
time $\tau_c$ is given by: $M_0\exp[-t/T_{2R}-t^2/2T_{2G}^2]$,
where $T_{2R}^{-1}=(\beta+R){}^{63}T_1^{-1}$ is the Redfield
term\cite{CPS}, and $T_{2G}$ is the Gaussian part of the echo
decay assuming static neighbors. Here $\beta=3$ and $R$ is the
$T_1$ anisotropy ratio.  Since $\tau_c$ is distributed,
${}^{63}T_1^{-1}$ is as well.  Intensity is determined by
extrapolating the spin echo decay data (available for $t\! > \!
 2 t_r$) to $t$ = 0; this extrapolation is dominated by the Redfield
term, so we ignore the Gaussian contribution in the following
discussion. The full signal from all of the nuclei is then given
by:
%\begin{equation}
\(M(t)=\int{M_0 \exp[-t/T_{2R}(\tau_c)]P(\tau_c)d\tau_c} \),
%\end{equation}
where $P(\tau_c)d\tau_c =P(E_a)dE_a$. For $T \gg E_0 / k_B$,
$M(t)$ decays exponentially with the single time constant
$T_2(\tau_c \! = \! \tau_\infty)$. For \( T \lesssim E_0/k_B \)
some regions relax rapidly causing $M(t)$ to drop sharply for
short $t$; single exponential behavior is then recovered at longer
$t$. One can show that the extrapolated intensity $I_0(T) = \int
_0 ^{\kappa k_BT} {P(E_a) dE_a}$, where \( \kappa \equiv \ln (\tau
_{\rm cut} / \tau_{\infty} )\). For \(\tau \! > \! \tau_{\rm
cut}\) excessive signal decay occurs for $t\ge 2t_r$, thus wiping
out the signal [we use the criterion \(M(2t_r;\tau_{\rm cut}) =
1\%\)]. The measured low temperature anisotropy of \ltoi
\cite{suhlesco} shows $H_c^2\simeq 0$ hence \(R \simeq
\frac{1}{2}\). Using Eq.\ \ref{eqn:t1vstau} and assuming $A_{\rm
hf}= 100$ kOe/$\mu$B (the wave vector dependence of the very slow
spin fluctuations is not known) gives \(\tau_{\rm cut} \! = \!
0.13\) ps \(\Rightarrow \kappa\! = \! 1.5\). In Fig.\
\ref{fig:cuint} we show a plot of $I_0(T)$ using this value of
$\kappa$ and the distribution $P(E_a)$ extracted from fits to the
La $T_1$ data; the agreement of this straightforward model with
the data is good, especially given the uncertainty in parameters.
This strongly indicates that Cu wipeout is not a measure of the
stripe order parameter\cite{hunt}.

The Cu spin lattice relaxation rate, \ctoi , was measured between
4 and 300 K. The inversion recovery data were fit to the standard
expression for magnetic relaxation of the central transition with
a single $T_1$\cite{relaxpaper}. The temperature dependence of
$(^{63}T_1T)^{-1}$ is shown in the inset of Fig.\ \ref{fig:lat1},
as well as data from Ohsugi {\it et al}.,\cite{ohsugi} for \lsco.
The spin lattice relaxation rate of the Cu that contribute to the
NMR signal at low temperatures reveals no unusual behavior, and is
in fact quite similar to the $(^{63}T_1T)^{-1}$ in \lsco. Note
that $(^{63}T_1T)^{-1}$ is identical for the two systems for
$T>50$ K. Below this the \lsco\ becomes superconducting at 38 K,
whereas the \lesco\ does not.

The character of the Cu echo decay (Fig.\ \ref{fig:cuint} inset)
also provides evidence for $T_1$ inhomogeneity. The temperature
dependence of the echo decay in the \lesco\ is similar to that in
other cuprates, except for a distinct crossover from Gaussian to
exponential decay around 100 K.  In spite of the distributed
Redfield contribution discussed above we would expect Gaussian
behavior at large $t$; this is not observed experimentally. This
behavior can be understood if the measured Cu nuclei are coupled
to neighboring spins that experience a fast spin lattice
relaxation. For like neighbors the echo decay is given by
$M(t)=M_0\exp[-t^2f(t/T_1)/2T_{2G}^2]$, where \(f(x) =
8x^{-2}[5x/2+9e^{-x/2}-2e^{-x}-7]\)\cite{curroJMR}. For infinite
$T_1$ the echo decay is Gaussian; as $T_1$ gets shorter, the echo
decay becomes exponential.  The solid lines through the data in
the inset of Fig.\ \ref{fig:cuint} are fits in which we assume
$T_1^{-1}={}^{63}T_{\rm 1,meas}^{-1}+{}^{63}T_{{\rm
1,extra}}^{-1}$, where the extra contribution is a variable
parameter that represents the contribution of fast relaxation on
the neighbors\cite{comment}. Qualitatively, these fits indicate
that the observed Cu nuclei are coupled to neighbors that undergo
$T_1$ fluctuations much faster than those of the observed nuclei.

Combined, the La and Cu data reveal an inhomogeneous, glassy
freezing of the spin dynamics in the heavily doped, stripe ordered
systems.  La relaxation demonstrates a dramatic slowing of spin
fluctuations such that their characteristic frequency matches
$\omega_L \sim 5$--50 MHz around 10 K, and the data are well
modeled by a distribution of activation energies. The quantitative
explanation for the loss of Cu intensity from spin freezing
provides a direct demonstration that the fluctuations are
inhomogeneous: at any given temperature $0 \! < T \! < 100$ K a
$T$-dependent fraction of the Cu spins experience spin
fluctuations so slow (see Eq.\ \ref{eqn:t1vstau}) that they are
``invisible'' while the remaining spins remain visible due to much
faster spin fluctuations.

We have seen that quantitatively similar inhomogeneous freezing of
spin fluctuations (that are well modeled by very similar
distributions of activation energies) occurs in a variety of hole
doped \lc s containing impurities of very different character: i)
lightly Sr-doped \lc\ containing only a few out-of-plane
impurities, ii) lightly Li-doped \lilco\ where the impurities are
in the \cuot\ planes, iii) both lightly and heavily Sr-doped
\lesco\ where Eu co-doping adds more out-of-plane impurities and
induces the LTT structure thought to pin charged stripes.  The
fact that the spin freezing is independent of both impurity
density and the location of the impurity (in or out of the plane
where the fluctuating moments and the charged stripes reside)
strongly suggests that extrinsic impurities are not an essential
element of the inhomogeneity, rather it appears to be an {\it
intrinsic} response of the hole doped system to perturbation.
Finally Hunt \etal. have shown a similar loss of Cu intensity in
\lscox\ throughout the doping range \( {{1} \over {16}} < x < {{1}
\over {8}}\) \cite{hunt} further emphasizing the generality of
this phenomenon. We note that loss of NMR signal has been observed
previously in classical spin glass systems\cite{CPSpaper} and in
stripe-ordered lanthanum nickelate\cite{brom:lno}.

Two different mechanisms have been explored in the literature in
order to explain the spin freezing in the antiferromagnetic state:
one involves the ``cluster spin glass'' and freezing of the
transverse spin degrees of freedom\cite{chou}; another the loss at
low temperatures of collective hole/stripe motion\cite{suhllco}.
In the presence of frustration a Heisenberg system exhibits glassy
behavior of the transverse spin components below the longitudinal
ordering temperature\cite{gooding1}, and Kivelson \etal.\ have
pointed out the role of disorder in striped
systems\cite{Kivelson:liqxtal}. The observed very slow spin
fluctuations imply large clusters of spins (presumably frustrated
by disorder) are involved. The role of charged stripes in these
dynamics isn't clear. Stripe motion is certainly a mechanism for
spin dynamics\cite{suhllco}, however if the same mechanism is
responsible for the inhomogeneous relaxation over the wide regime
of hole density we observe, then the very weak hole density (hence
stripe density) dependence of the distribution of activation
energies implies stripe-stripe interactions play no role over a
decade of stripe density; this seems unlikely. Because the stripes
constitute anti-phase domain walls in the antiferromagnet, stripe
disorder leads to severe spin disorder\cite{swc:NSlsno,yosh:lsno}.
Hasselmann and Castro-Neto have pointed out that for $x \sim 1/8$
the frustration can arise from point defects and disorder in the
stripe topology\cite{hasselmann}. The slow inhomogeneous dynamics
could arise from frustrated interactions between such large
clusters.

The authors thank N.\ Hasselmann, A.\ Castro-Neto and S.\ Kivelson
for valuable discussions.  The work at Los Alamos was performed
under the auspices of the US Department of Energy. The work at
University of K\"oln was supported by the Deutsche
Forschungsgemeinschaft through SFB 341.  M.H.\ acknowledges
support by the Graduiertenstipendium des Landes
Nordrhein-Westfalen.

%\vspace{-3mm}

%\end{multicols}

\begin{thebibliography}{29}
\bibitem{comment}  One must also account for the fraction
of unlike $^{65}$Cu isotope; see Ref.
\protect\onlinecite{curroJMR}.
\end{thebibliography}
\end{document}